\documentstyle[aps]{revtex}
\begin{document}

\title{  
      On the nucleon-nucleon interaction leading to 
      a standing wave instability in symmetric nuclear matter
       }

\author{Janusz Skalski }

\address{
 So\l tan Institute for Nuclear Studies,\\
ul. Ho\.za 69, PL- 00 681, Warsaw, Poland \\
 e-mail: jskalski@fuw.edu.pl
 }

\date{\today}

\maketitle

\begin{abstract}
  We examine a recently proposed nucleon-nucleon interaction, claimed by its authors 
  both realistic and leading to a standing wave instability in symmetric nuclear matter.  
  Contrary to these claims, we find that this interaction   
   leads to a serious overbinding of $^4$He, $^{16}$O and $^{40}$Ca nuclei when 
   the Hartree-Fock method is properly applied. The resulting nuclear densities 
  contradict the experimental data and all realistic Hartree-Fock results. 
  \end{abstract}

\pacs{PACS number(s):  }

\section{Introduction}
 
  Recently, a simple nucleon-nucleon interaction was proposed which is  
  claimed both realistic and leading to a standing wave instability in  
  symmetric nuclear matter \cite{Over}. Although, strictly speaking, 
  symmetric nuclear matter is a purely speculative object, it served for
  years as a testing ground for nuclear many-body theories and new insights into its 
  properties are of considerable interest. The hint that 
  all these theories missed the spatial modulation of the nuclear matter 
  density is provocative. When putting forward such claim, one has to make sure  
  that the proposed interaction satisfies constraints imposed by our knowledge of 
  nuclear physics.  
  
  The interaction considered in \cite{Over} reads  
  \begin{equation}
   \label{inter}
   V({\bf r}_1,{\bf r}_2)= - \alpha C  ({\bf r}_1 - {\bf r}_2)^2 
  e^{-({\bf r}_1 - {\bf r}_2)^2/s^2} + 
  \beta \sqrt{<T>} \delta({\bf r}_1 - {\bf r}_2) ,
  \end{equation}
   where $<T>$ is the center-of-mass corrected average kinetic energy:
 \begin{equation}
 <T>=(\frac{1-1/A}{A}) \frac{\hbar^2}{2m} \sum_{i=1}^{A} \mid \nabla \phi_i \mid^2 . 
 \end{equation} 
 In the latter equation, single-particle orbitals $\phi_i$ relevant for the Hartree-Fock 
 (HF) treatment are explicitly introduced.
 The auxiliary constant $C=2 \pi^{-3/2}s^{-5}/3$, while the strength and range 
 of attraction and the strength of contact repulsion are chosen as 
 $\alpha=$1690 MeV fm$^3$, $s=0.54$ fm, $\beta=225$ MeV \cite{Over}. 
 These parameters were intended to fit 
 the binding energy and the equilibrium density of nuclear matter and the 
 binding energy of alpha particle (but see below). 
 With these parameters, the authors reported reasonable 
  values of the compressibility modulus of nuclear matter and of binding energies of 
  even-even $N=Z$ nuclei. 

  It is crucial to understand that, although the authors refer to the HF method when 
  describing their calculations for finite nuclei \cite{Over}, they performed in fact 
 only a very restricted minimization. This restriction is evident in a very 
 small harmonic oscillator basis that has been used.   
 In addition, the exchange integrals were not calculated, but assumed to be a 
  fraction of the direct terms, depending on the average kinetic energy. 
 Therefore, especially in view of quite important  consequences claimed,
 an independent evaluation of the results of Ref.\cite{Over} is called for. 

  In this short note we report the results of regular HF calculations with the interaction (\ref{inter})  
  by which we determined binding energies of $^4$He, $^{16}$O and $^{40}$Ca 
  nuclei. These binding energies are at variance with \cite{Over} and the obtained matter and charge 
  densities of $^{16}$O and $^{40}$Ca are highly unusual.

\section{Hartree-Fock method}

As the interaction (\ref{inter}) is spin- and isospin independent one 
assumes the fourfold degeneracy of the HF orbitals for even-even $N=Z$ nuclei of interest. 
 There are $A/4$ independent orbitals and we 
 sum over them to obtain density $\rho=\sum_{i=1}^{A/4} \mid \phi_i \mid^2 $.
 Neutron and proton densities are equal, $\rho_n=\rho_p =2\rho$, and the total density equals $4\rho$. 
 Similarly, the total average kinetic energy $<T>$ is quadruple of the sum of 
 kinetic terms over independent orbitals.  

 The HF energy reads 
\begin{eqnarray}
\label{E}
\lefteqn{E = <T> + 8\int\int d^3r_1d^3r_2\rho({\bf r}_1) \rho({\bf r}_2) V_a({\bf r}_1 - {\bf r}_2) }
    \nonumber \\
 & & -2 \sum_{i,j}^{A/4} \int\int d^3r_1 d^3r_2 \phi_i^*({\bf r}_1)\phi_j^*({\bf r}_2)
  \phi_j({\bf r}_1)\phi_i({\bf r}_2)V_a({\bf r}_1 - {\bf r}_2) + 6\beta<T>^{1/2}\int d^3r \rho^2 ,
\end{eqnarray} 
 where $V_a$ is the attractive part of (\ref{inter}). 
 Remembering about differentiation of $<T>$, 
 we obtain from (\ref{E}) a set of HF equations for 
 the wave functions $\phi_i$ and single-particle energies $\epsilon_i$
\begin{eqnarray}
\label{hf}
\lefteqn{ -\frac{\hbar^2}{2m^*} \nabla^2 \phi_i +3\beta \sqrt{<T>}\rho \phi_i 
 +4\int d^3r_2 \rho({\bf r}_2)V_a({\bf r}_1 - {\bf r}_2) \phi_i } \nonumber \\
 & & -\sum_{j}^{A/4} \int d^3r_2 \phi_j^*({\bf r}_2)
  \phi_j({\bf r}_1)\phi_i({\bf r}_2)V_a({\bf r}_1 - {\bf r}_2) 
 = \epsilon_i \phi_i    ,
\end{eqnarray}
where the effective mass is given by $m/m^*=(1-1/A)
[1+3\beta I/(A\sqrt{<T>})]$, with $I=\int d^3r\rho^2$.

 The numerical solution for $\phi_i$ is straightforward, but tedious due to the 
 exchange integrals. For spherically symmetric completely filled shells 
 there is, however, a well known Slater method \cite{Slater} to obtain 
 the exact exchange potential. Adapting this general argument to the attractive potential 
 $V_a$ we can express the exchange potential acting on the wave function 
 $\phi_{n',l',m'}(r,\theta,\varphi)=R_{n'l'}(r) Y_{l'm'} (\theta,\varphi)$ as 
\begin{eqnarray}
\label{Expot}
 \lefteqn{V_{a  Ex}  \phi_{n'l'm'} =}  \nonumber \\
 & &  Y_{l'm'} \sum_{nl} R_{nl}(r) \left\{
 \sum_{k=\mid l-l' \mid}^{l+l'} \frac{2l+1}{2l'+1} A(k,l,l') 
 \int_0^{\infty} dr_2 r_2^2 R_{nl}(r_2)R_{n'l'}(r_2) V_k(r_1,r_2) 
 \right\}  ,
\end{eqnarray}
 where the subshell index $(nl)$ in the summation runs over the 
 occupied orbitals.   
 The coefficients $A(k,l,l')$ are given by 
\begin{equation}
\label{A}
 A(k,l,n)=\frac{2n+1}{2}\int_{-1}^{1} P_k P_l P_n ,
\end{equation}
where $P_i$ are Legendre polynomials. The functions $V_k$ define the expansion 
 of $V_a$ into spherical harmonics 
\begin{eqnarray}
\lefteqn{ V_a({\bf r}_1,{\bf r}_2)= \sum_k V_k(r_1,r_2) P_k(cos\theta) ,} & \\
 & &  V_k(r_1,r_2)= -\alpha C s^2 e^{-(r_1^2+r_2^2)/s^2} (2k+1) [(k+1+
 \frac{r_1^2+r_2^2}{s^2}) f_k(z)-zf_{k-1}(z)] ,  
\end{eqnarray}
 where $f_k$ are the spherical Bessel functions of the imaginary argument, i.e. 
 $f_k(z)=(-i)^kj_k(iz)$ = $\sqrt{\pi/(2z)}I_{k+1/2}(z)$, with $I_{k+1/2}(z)$ the modified Bessel function 
 \cite{Abr} and $z=2r_1r_2/s^2$.

\section{Results and discussion}

In order to have a check on the HF results we used two different schemes, one 
 spherical, using decomposition (\ref{Expot}), and the other three-dimensional. 
 Both use wave functions defined over a spatial mesh. The general three-dimensional scheme 
 being more time-consuming practically restricts the mesh size to about 30 points in each 
 direction in one octant of space. 
 The spherical scheme allows radial meshes of 100 points or more and may easily produce 
  accurate solutions. 

 We consider magic $^4$He, $^{16}$O and $^{40}$Ca nuclei for which the spherically symmetric solutions 
 are expected. The HF problem was solved by the imaginary-time evolution. 
 The convergence is rather slow for density and single particle energies, especially for $^{40}$Ca. 
 This is due to the buildup of central density peak which costs little energy in final stages of 
 iteration. 
 The three-dimensional scheme becomes impractical in this case but still its results tend to those 
  of the spherical code.  
  Below, we report densities calculted with the faster 
   spherical code on the mesh of 100 points.  
 
 As the starting wave functions we took the results of \cite{Over}. Therefore we could compare 
our initial energies and densities with those of \cite{Over}.  
 We obtain for $^4$He the same energy as in \cite{Over}, but for $^{16}$O and $^{40}$Ca we find 
  differences. These must be attributed to the error in energy introduced in \cite{Over} by the 
  approximate treatment of the exchange integrals. Indeed, the  
 integrals calculated analytically for $s$ and $p$ wave functions in $^{16}$O agree exactly with 
 the results of our numerical codes. 
  The correct values of the binding energy per nucleon for initial configurations  
  are 8.801 MeV in $^{16}$O and 11.137 MeV in $^{40}$Ca, to be 
  compared to 8.59 MeV and 10.76 MeV reported in \cite{Over}. Thus, the exact calculation of the 
  exchange integrals alone points to the overbinding problem with the interaction (\ref{inter}). 
  This problem is magnified if one cares for the HF solutions. 

 The first issue is the binding energy of alpha particle which bears on the determination of interaction 
 constants \cite{Over}.  The optimal wave function is more peaked than the gaussian used there. 
  The HF binding energy per nucleon is 8.076 MeV, i.e. 0.78 MeV more than in \cite{Over} 
  where the experimental value corrected for Coulomb interaction was used. 
  Thus, the interaction (\ref{inter}) overbinds already $^4$He when properly treated. 

 The results for three nuclei are collected in Table 1. As seen there, the overbinding of $^{16}$O, and 
  especially $^{40}$Ca, is really serious. For $^{16}$O, the calculated binding is 173.57 MeV without Coulomb 
 while the experimental value is 127.619 MeV \cite{Wap}. Allowing about 13 MeV for Coulomb energy 
 (the direct term minus exchange, as it results from any realistic HF) 
 we get more than 30 MeV overbinding.  For $^{40}$Ca, the calculated binding of 565.17 MeV, 
  even after subtraction of about 71 MeV of Coulomb repulsion, is by about 152 MeV (!) larger than the 
  experimental value of 342.052 MeV \cite{Wap}. In Table 1, we also give the difference in total binding 
  (without Coulomb) between our results and those of \cite{Over} to emphasise the importance 
   of proper HF minimization. 

 The calculated self-consistent nuclear densities are depicted in Figure 1. The tendency towards 
  central peak development is evident. It is worth noting that this tendency is already seen in 
  inaccurate results of \cite{Over}. The density of the $^{40}$Ca shown there exhibits a strange pile-up in the 
   center. However, the HF results shown in Figure 1 allow to appreciate that this problem is even more grave:  
  The central density is more than 1.5 times larger in the case of $^{16}$O, and three times larger in $^{40}$Ca 
  than the experimental one (see e.g. \cite{Bat}). 

\section{Conclusions}

The exact HF calculations with the recently proposed interaction (\ref{inter}) for magic $^4$He, $^{16}$O and 
$^{40}$Ca nuclei show a serious overbindig problem. Associated nuclear densities develop central 
 peaks taking a form unknown in nuclear physics. Both deficiencies grow with increasing mass.  
In view of the above results it is clear that the interaction proposed in \cite{Over} is very far 
 from a realistic nucleon-nucleon force. Therefore, the assertions about the standing wave instability in 
 nuclear matter made there, as related to unrealistic interaction, are unfounded.

\newpage 

\begin{center}
 
\hspace{5mm}{TABLE 1 - Calculated HF binding energies per nucleon vs. results of  
  \cite{Over} and the difference between total quantities (in [MeV]).   }

\vspace{9mm}

\begin{tabular}{c|ccc} 
\hline

   & $B/A$  & $B/A$ in \cite{Over}  &  $B-B$\cite{Over} \\
 \hline

 $^4$He  &  8.08 &  7.3  &   3.11  \\

 $^{16}$O &  10.85  & 8.59  &  36.13   \\

 $^{40}$Ca &  14.13  &  10.76  & 134.77   \\
 \hline
\end{tabular}

\vspace{3cm}

Figure caption 

\vspace{9mm}

Figure 1

\vspace{3mm}

 Total nuclear HF densities (thick lines) and densities from \cite{Over} (thin lines) for 
  $^{16}$O (dashed) and $^{40}$Ca (solid).

\end{center}

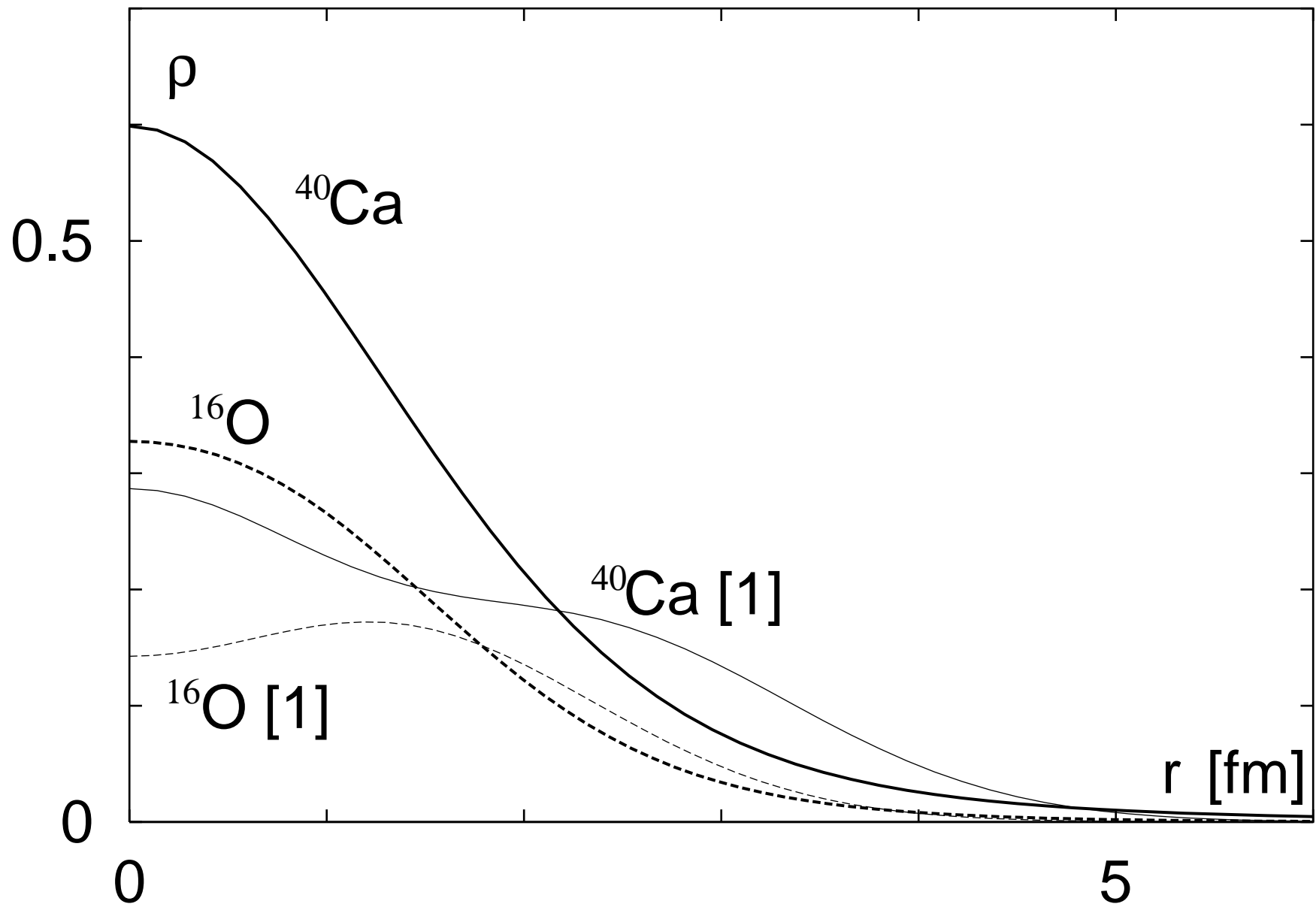
\end{document}